\documentclass[aps,twocolumn,pre,showpacs,eqsecnum]{revtex4-1}
\usepackage{graphpap}
\usepackage[dvips]{graphicx}
\usepackage[dvips]{graphics}
\usepackage{color}
\usepackage[normalem]{ulem} %emphasize weiterhin kursiv
\usepackage {ulem} %\emph{Text}: Text wird unterstrichen

\begin{document}

\title{Fabrication of coupled graphene-nanotube quantum devices}
 \author{S. Engels$^{1,2,3}$, P. Weber$^{1,2,3}$, B. Terr\'{e}s$^{1,2,3}$, 
 J. Dauber$^{1,2,3}$, C. Meyer$^{2,3}$, C. Volk$^{1,2,3}$, S. Trellenkamp$^{2}$, U. Wichmann$^{1}$ and C. Stampfer$^{1,2,3}$}
\email{stephan.engels@rwth-aachen.de}
\affiliation{ 
$^1$II. Institute of Physics B, RWTH Aachen University, 52074 Aachen, Germany, EU\\
$^2$Peter Gr\"unberg Institute (PGI-6/8/9), Forschungszentrum J\"ulich, 52425 J\"ulich, Germany, EU\\
$^3$JARA -- Fundamentals of Future Information Technologies
}

\date{ \today}

 \begin{abstract}
We report on the fabrication and characterization of all-carbon hybrid quantum
devices based on graphene and single-walled carbon nanotubes. We discuss
both, carbon nanotube quantum dot devices with graphene charge detectors and
nanotube quantum dots with graphene leads. The devices are fabricated by
chemical vapor deposition growth of carbon nanotubes and subsequent
structuring of mechanically exfoliated graphene. We study the detection of individual
charging events in the carbon nanotube quantum dot by a nearby graphene nanoribbon and show that they lead
to changes of up to 20$\%$ of the conductance maxima in the graphene nanoribbon acting as a
good performing charge detector. Moreover, we discuss an electrically coupled
graphene-nanotube junction, which exhibits a tunneling barrier with tunneling rates
in the low GHz regime. This allows to observe Coulomb blockade on a carbon nanotube quantum dot with graphene source and drain leads.
\end{abstract}

 %\vspace{0.1cm}
 \keywords{graphene, carbon nanotube, charge detector, quantum dot}
 \pacs{71.10.Pm, 73.21.-b, 81.07.Ta, 81.05.ue}
 \maketitle

\newpage
Carbon nanomaterials, such as graphene and carbon nanotubes (CNTs) attract increasing interest 
mainly due to their promises
for flexible electronics,  high-frequency devices and spin-based quantum circuits \cite{lee11,wu2011,Lin2010}.
Both materials consist of sp$^2$-bound carbon and exhibit unique electronic properties resulting in the 
suppression of direct backscattering, high
carrier mobilities and low intrinsic spin noise.
In particular the weak hyperfine interaction makes graphene and CNTs interesting host materials for 
quantum dots which promise the implementation of long-living spin qubits \cite{tra07}.
Up to the present state quantum dots (QDs) and double quantum dots have been demonstrated successfully in both carbon nanomaterials.
In particular, the fabrication of ultra-clean few-carrier QDs with well-defined spin states in quasi one-dimensional (1-D) carbon nanotubes attracted great interest \cite{sap05,chu09,jor08}.
A comparable quality in graphene is not yet reached, which is mainly due to its gap less band structure making it hard to controllably confine electrons and holes. State-of-the-art graphene QDs are therefore either based on nanoribbons \cite{liu09,liu10} or etched islands \cite{sta08a,sta08b,pon08,mol09a,vol11} and in both cases edge roughness and disorder are dominating their properties. However, in contrast to quasi 1-D nanotubes, the 2-D nature of graphene makes it easy to integrate lateral graphene gates and in-plane charge sensors \cite{gue08}, which are both important for the control and readout of QD states.
Here, we present the fabrication and characterization of quantum devices based on both graphene and carbon nanotubes, which combine the advantages of the two carbon allotropes and open the route to unprecedented quantum devices.
\begin{figure}[hbt]\centering
\includegraphics[draft=false,keepaspectratio=true,clip,%
                   width=1\linewidth]%
                   {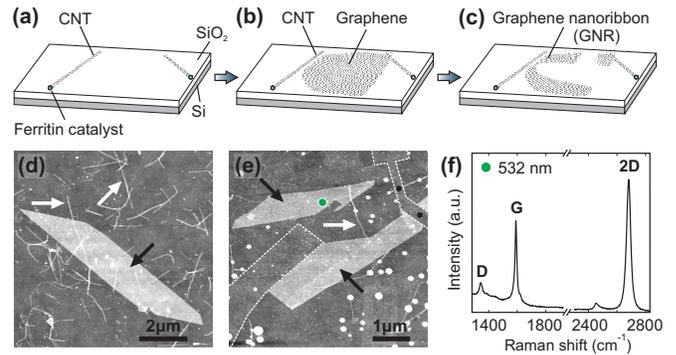}
\caption[FIG1]{(color online) (a),(b) and (c) Schematic illustrations of the three main fabrication steps. (d) Scanning force microscopy (SFM) image of a $\rm{SiO}_2$ substrate with CVD grown CNTs (white arrows) and subsequently deposited graphene (black arrows). (e) SFM image of a similar sample after the dry etching process. White dashed lines indicate areas exposed to an $\rm{Ar/O}_{2}$ plasma which results in etching both carbon materials as indicated by the black points. (f) Raman spectrum of the graphene flake at the position indicated by the green dot in panel (e). The spectrum shows clear characteristics of single-layer graphene.}
\label{fig1}
\end{figure}
\newline
In particular, we first discuss a carbon nanotube QD with a capacitively coupled graphene nanoribbon acting as electrostatic gate and charge detector. Charge read-out schemes have proven to be difficult for carbon nanotubes QDs, the main challenge lying in the random orientation and position of the nanotubes on the substrate. Two possible strategies have been put forward so far to overcome this difficulty. The first one consists in placing a metallic single electron transistor (SET) close the the CNT QD \cite{got08,bie06}. The second relies in capacitively coupling the QD to an SET realized in the same \cite{chu09} or in a neighboring CNT \cite{zho12} via a deposited metallic gate. Our work adds a third detection scheme based on a graphene charge detector which is easy to fabricate and has a more than sufficient charge sensitivity to detect single charging events.
In addition to the charge sensing, we show a graphene-carbon nanotube hybrid device where a nanotube QD is contacted with both source and drain graphene leads.
Thus, a quantum dot device exclusively built of the two different carbon allotropes. This configuration is especially interesting since it opens the route to quantum dot devices with a tunable density of states in the source and drain leads. In particular, it may allow to experimentally investigate the pseudo-gap Kondo model which predicts that graphene undergoes a quantum phase transition \cite{sac00, gon98} from a phase with a screened to an unscreened impurity moment at low charge carrier densities \cite{voj10}. The predicted quantum phase transition might manifest itself in the characteristic scaling of observable quantities (e.g. the Kondo resonance) in a graphene-contacted quantum dot.
\begin{figure}[tb]\centering
\includegraphics[draft=false,keepaspectratio=true,clip,%
                   width=1\linewidth]%
                   {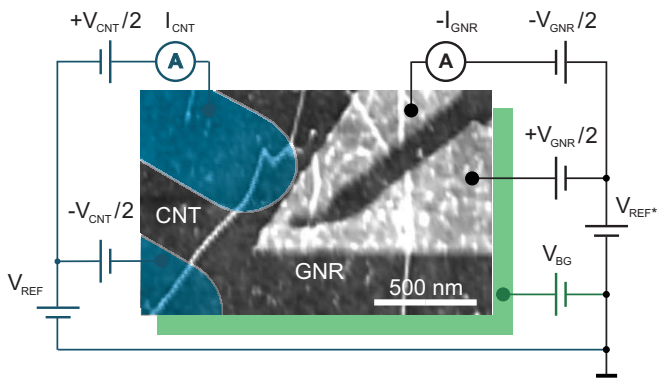}
\caption[FIG2]{(color online) Scanning force micrograph of 
a carbon nanotube (CNT) quantum dot device with a nearby structured
graphene nanoribbon (GNR) for charge sensing. The schematic illustration
shows the used measurement circuit and the applied voltages. For detailed
information see text.
}
\label{fig2}
\end{figure}
\newline
The fabrication process is based on chemical vapor deposition (CVD) growth of carbon nanotubes and subsequent deposition of
mechanically exfoliated natural graphite. In Figs.~\ref{fig1}(a) to \ref{fig1}(c) we show the three main fabrication steps for making all-carbon graphene-nanotube devices. 
As a first step single-walled carbon nanotubes are grown on 290 nm $\rm{SiO}_{2}$ on highly p-doped Si substrates by CVD using a Ferritin-based iron catalyst method \cite{dur08}. 
The single-walled carbon nanotubes have a diameter of around 1.5-2~nm and are up to several micrometers in length. %\newline
In a next step, graphene is deposited on these pretreated substrates by mechanical exfoliation of natural graphite~\cite{nov04}. Finally, a graphene nanoribbon (GNR) or graphene leads are patterned from the deposited graphene by an electron beam lithography (EBL) step followed dry etching.
Figs.~\ref{fig1}(d) and \ref{fig1}(e) show scanning force microscopy (SFM) images of two examples of the deposition process.
Due to the stochastic nature of the method a great variety of different configurations is obtained. For example, the graphene flake shown in Fig.~\ref{fig1}(d) (black arrow) lies on top or nearby of a number of individual CNTs (white arrows) whereas in Fig.~\ref{fig1}(e) we show an example where a nanotube (white arrow) is contacted by two graphene flakes (black arrows).
A crucial parameter for the success of this stochastic fabrication process is the density of grown CNTs which can be well controlled by the Ferritin-based CVD process \cite{dur08}. In this study we used a density of approx. 1-2 CNTs per $\rm{\mu m}^2$. 
In Fig.~\ref{fig1}(f) we show a Raman spectrum taken at the position indicated by the green dot in Fig.~\ref{fig1}(e). From the full width at half maximum of the 2D peak of 34 $\rm{cm}^{-1}$ and the relative intensity of the 2D and G peak of I(G/2D)$\approx$0.65, we conclude that the graphene is of single-layer nature \cite{fer06,gra07} (laser excitation of 532 nm). A Raman spectrum of the bottom graphene flake at the position of the intersecting CNT reveals a bilayer flake (not shown).
The contacted CNT structure consists of two segments. Most likely the upper segment is a small bundle of single-walled CNTs whereas the lower segment (white arrow in Fig. \ref{fig1}(e)) is an individual  single-walled CNT. This conclusion is supported by a comparison of the SFM profiles of the single-layer graphene and the lower carbon nanotube segment, as well as the high accuracy of the Ferritin-based CVD growth process \cite{dur08}.
The third fabrication step consists of EBL followed by an $\rm{Ar/O}_{2}$ based dry etching step for (i) structuring graphene and (ii) removing unwanted CNTs. In Fig. \ref{fig1}(e) we highlight areas (dashed lines) where graphene and CNTs have been successfully removed (see black points).
Finally, we used EBL, metal evaporation (5~nm Cr/ 50~nm Au) and lift-off for contacting the devices.
\begin{figure}[tb]\centering
\includegraphics[draft=false,keepaspectratio=true,clip,%
                   width=1\linewidth]%
                   {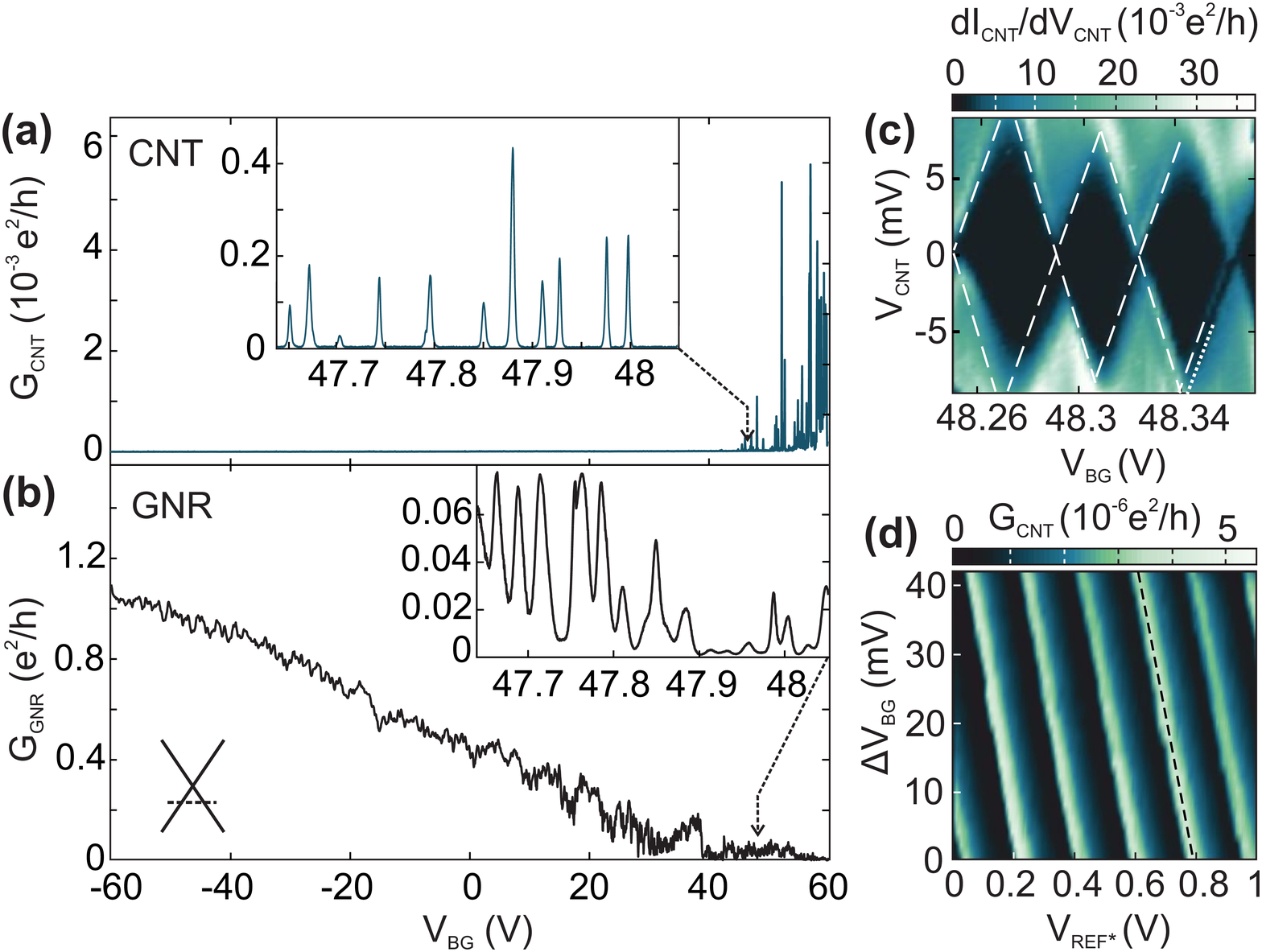}
\caption[FIG3]{(color online) (a),(b) Back gate characteristics of the carbon nanotube QD (a) and the graphene nanoribbon (b). Both measurements were recorded at a source-drain bias voltage of $\rm{V}_{CNT}$=$\rm{V}_{GNR}$=0.5~mV and $\rm{V}_{REF}$=0~V. The upper inset in panel (a) shows Coulomb blockade resonances of the CNT QD as a function of $\rm{V}_{BG}$. The inset in panel (b) highlights the conductance resonances in the GNR in the same range of $\rm{V}_{BG}$. (c) Differential conductance on the CNT QD. The white dotted line indicates an excited state with an energy of $\Delta$=0.9~meV. (d) Conductance of the CNT as function of $\Delta \rm{V}_{BG}$=$\rm{V}_{BG}$-13.63~V and $\rm{V}_{REF^{*}}$ at $\rm{V}_{CNT}$=20~mV. The black dashed line highlights the relative lever arm $\rm{\alpha}_{BG,GNR}$.
}
\label{fig3}
\end{figure}
\begin{figure*}[t]\centering
\includegraphics[draft=false,keepaspectratio=true,clip,%
                   width=0.8\linewidth]%
                   {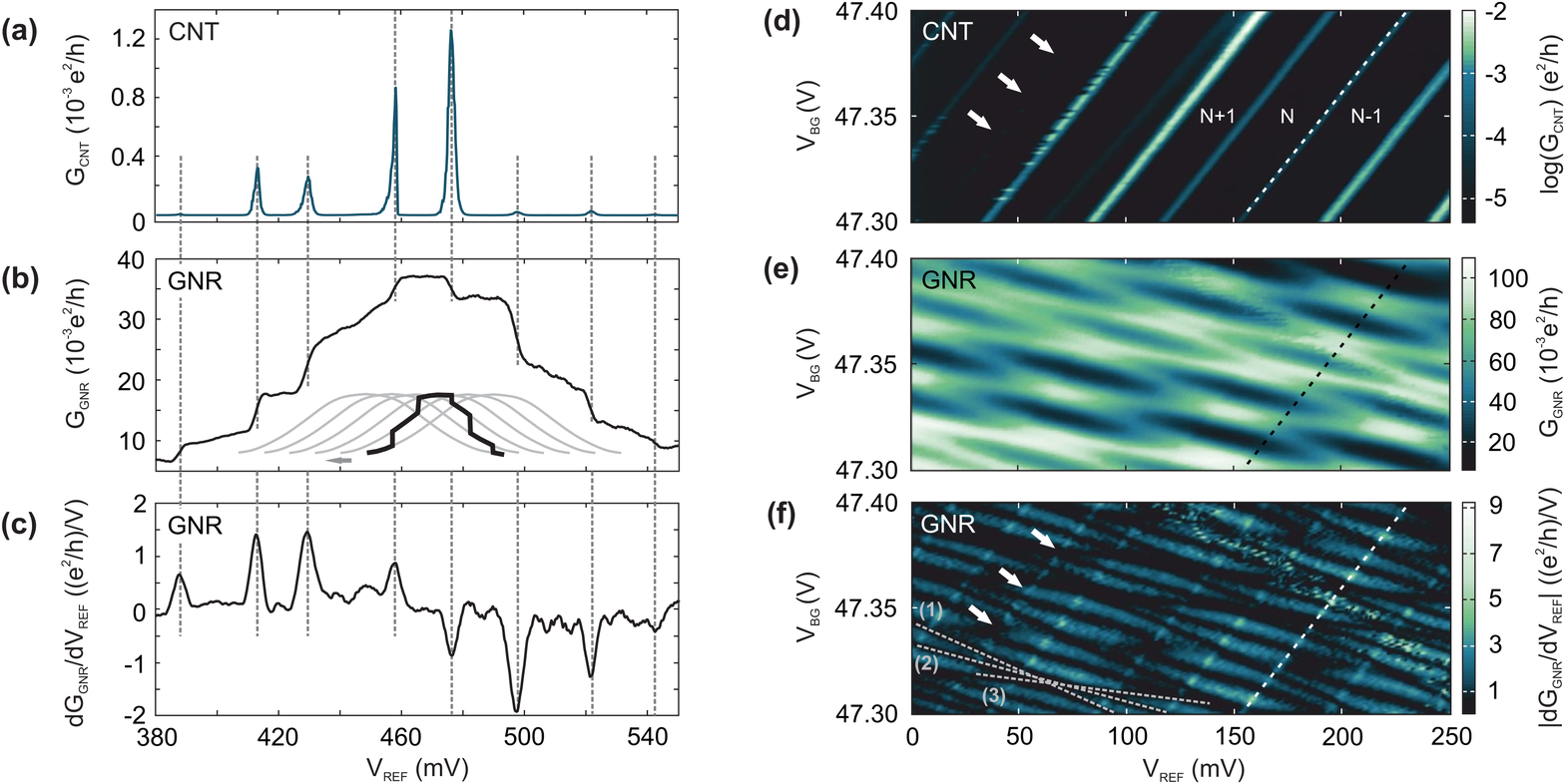}
\caption[FIG4]{(color online) (a) QD conductance and (b) GNR conductance as a function of $\rm{V}_{REF}$ with a simultaneous change in BG voltage following $\rm{V}_{BG}$=a-b$\cdot$$\rm{V}_{REF}$ with a=48.166 V and b=-0.335. A bias voltage of $\rm{V}_{CNT}$=$\rm{V}_{GNR}$=$0.5$~mV was applied to both structures. The inset in (b) shows a schematic of the charge detecting mechanism. (c) Derivative of (b) showing well pronounced peaks at every charging event.
(d) Dependence of the CNT QD conductance on $\rm{V}_{BG}$ and $\rm{V}_{REF}$ measured at $\rm{V}_{CNT}$=0.5 mV on a logarithmic scale. Periodic Coulomb blockade resonances with a positive slope of $\rm{\alpha}_{REF,BG}=$1.25 are observed. The resulting electron occupation numbers of the QD are given in white letters.
(e)~Conductance of the GNR depending on $\rm{V}_{BG}$ and $\rm{V}_{REF}$ with $\rm{V}_{GNR}$=0.5 mV.
(f) Absolute value of the GNR transconductance. Elevated conductance traces with a positive slope (dotted line) match perfectly with the Coulomb resonances in (d). Detection lines are even visible where the current is too small to be measured directly (see white arrows). Measurements shown in (a)-(c) were measured parallel to line (2). All measurements were recorded simultaneously.
}
\label{fig4}
\end{figure*}
\newline
Fig.~\ref{fig2} shows a SFM image of an all-carbon device consisting of a carbon nanotube lying in the close vicinity to an etched graphene nanoribbon (GNR) which acts as a charge detector (CD). By Raman spectroscopy we identify the bright colored area to be a graphene nanoribbon of bilayer nature (not shown) and following the above argument we conclude that the nanotube is a single-walled carbon nanotube. Both carbon nanostructures are separated by roughly 150~nm, the nanoribbon has a width of around 100~nm and the CNT quantum dot is defined by two metal electrodes (indicated in blue) which are separated by 350~nm.
As illustrated in Fig.~\ref{fig2}, we apply a symmetric bias voltage ($\rm{V}_{CNT}$ and $\rm{V}_{GNR}$ respectively) to both structures. The overall Fermi level can be tuned by the back gate voltage $\rm{V}_{BG}$ applied to the highly doped Si substrate. Additionally, we can use the CNT (GNR) as a lateral gate for the GNR (CNT) by applying a reference potential $\rm{V}_{REF}$ ($\rm{V}_{REF^{*}}$).
All presented measurements were performed in a pumped $^4\rm{He}$-cryostat at a base temperature of T$\approx$1.5~K using low-frequency lock-in techniques.\newline
Figs.~\ref{fig3}(a) and \ref{fig3}(b) show the back gate characteristics i.e. the conductance as function of $\rm{V}_{BG}$ of both, the CNT (Fig.~\ref{fig3}(a)) and the graphene nanoribbon (Fig.~\ref{fig3}(b)) for a constant bias voltage $\rm{V}_{CNT}$=$\rm{V}_{GNR}$=0.5~mV. The CNT reveals a semiconducting behavior with a large band gap resulting in an extended BG region of suppressed current (-60~V$< \rm{V}_{BG} < $40~V). A high resolution measurement performed at the edge of this gap (see inset in Fig.~\ref{fig3}(a)) exhibits reproducible, well resolved and sharp Coulomb peaks. The peak width of the sharpest resonances is given by the electron temperature \cite{bee91} which can be extracted to be 2~K. In Fig.~\ref{fig3}(c) we show so-called Coulomb diamond measurements i.e. the differential conductance d$\rm{I}_{CNT}$/d$\rm{V}_{CNT}$ plotted as function of $\rm{V}_{BG}$ and $\rm{V}_{CNT}$. From the extent of suppressed current of the diamonds in bias ($\rm{V}_{CNT}$) direction we can extract a charging energy of $\rm{E}_C \approx$6-9 meV of the CNT quantum dot and a BG lever arm, $\rm{\alpha}_{BG}$=0.22. Following Ref. \cite{boc97} we can relate the charging energy with the length of the nanotube segment forming the quantum dot ($\rm{L}_{QD}$) by roughly $\rm{E}_C$=1.4 eV/$\rm{L}_{QD}$(nm). This provides an order of magnitude estimate of $\rm{L}_{QD}\approx$150-235nm which is in reasonable agreement with the device geometry.
The change in differential conductance parallel to the diamond edges is attributed to excited states providing additional transport channels. The observed excited states exhibit an energy of $\Delta$$\approx$0.9~meV (see e.g. white dotted line in Fig.~\ref{fig3}(c)).
\newline
The independently measured back gate characteristic of the bilayer graphene nanoribbon is shown in Fig.~\ref{fig3}(b). This low bias ($\rm{V}_{GNR}$=0.5~mV) measurement highlights that transport can be tuned from the hole regime (left inset) into the so-called transport gap \cite{che07, han07, mol09, sta09, tod09, gal10, han10, ter11} starting at around $\rm{V}_{BG}$$\approx$40~V. Within the gap region transport is governed by localized states resulting in sharp resonances of the conductance as shown in the right inset of Fig.~\ref{fig3}(c). Interestingly, both carbon nanostructures have different doping levels. In contrast to the n-doped CNT we observe a significant p-doping of the GNR which is most likely due to atmospheric $\rm{O}_{2}$ binding on the graphene edges \cite{ryu10}. Fig.~\ref{fig3}(d) shows the conductance of the CNT in dependence of the BG voltage $\rm{V}_{BG}$=13.63~V + $\Delta \rm{V}_{BG}$  and the reference voltage on the nanoribbon $\rm{V}_{REF^{*}}$ at $\rm{V}_{CNT}$=20~mV, which demonstrates the gating effect of the GNR on the CNT QD. Lines of higher conductance can be attributed to Coulomb resonances of the CNT and from their slopes we extract a relative lever arm of $\rm{\alpha}_{BG,GNR}$=0.23.\newline
If both quantum devices are now operated simultaneously, the nanoribbon device can be used to detect individual charging events on the nanotube QD device. This phenomenon is shown in Fig. 4. The BG voltage is put to an offset ($\rm{V}_{BG}$=48.166~V) such that (i) the CNT is in the Coulomb blockade regime and (ii) the conductance of the graphene nanoribbon exhibits sharp and well-reproducible resonances. Figs.~\ref{fig4}(a) and \ref{fig4}(b) show the simultaneously measured low-bias ($\rm{V}_{CNT}$=0.5~meV) conductance through the nanotube $\rm{G}_{CNT}$ and the nanoribbon $\rm{G}_{GNR}$ as function of $\rm{V}_{REF}$ where the BG voltage is simultaneously adjusted according to $\rm{V}_{BG}$=a-b$\cdot$$\rm{V}_{REF}$ with a=48.166 V and b=-0.335.
Similar to the inset of Fig.~\ref{fig3}(a), $\rm{G}_{CNT}$ exhibits Coulomb peaks, which are indicating single charging events in the CNT QD. In the simultaneously measured trace of the GNR, we observe distinct steps in the conductance at the exact positions of the CNT QD charging events. The steps in conductance can measure up to 20~$\%$ of the total resonance amplitude and are due to the capacitive coupling of both nanostructures. Increasing the reference potential $\rm{V}_{REF}$ and decreasing $\rm{V}_{BG}$ both leads to a higher chemical potential in the QD and subsequently to a lower occupation number at every event. Consequently, the GNR resonance shifts to lower values of $\rm{V}_{REF}$ giving rise to the shape of the charge detecting resonance in Fig.~\ref{fig4}(b). The unconventional shape of the detection signal is explained by the inset of Fig.~\ref{fig4}(b). In this schematic, each resonance curve (in gray) would represent the conductance resonance of the GNR at a \textit{fixed} charge state of the CNT QD. However, the latter is itself a (discontinuous) function of $\rm{V}_{REF}$, so that the measured conductance through the GNR is not a smooth curve but show a step whenever the charge state of the CNT QD (and thus the conductance resonance) changes. By relating the step height to the noise level of our measurement system we achieve an estimate for the charge sensitivity with an upper limit of $~10^{-3} \rm{e}/\sqrt{\rm{Hz}}$ which is in agreement with previous experiments on GNR charge detectors \cite{gue08}. In order to highlight the charge detection we further plot the transconductance i.e. the derivative of $\rm{G}_{GNR}$ with respect to $\rm{V}_{REF}$ as shown in Fig.~\ref{fig4}(c). Each individual transconductance peak (dip) is very well aligned with the directly measured Coulomb peaks. For proving this more rigerously we performed measurements as function of both $\rm{V}_{REF}$ and $\rm{V}_{BG}$ and extracted relative lever arms. Corresponding charge stability diagrams are shown in Figs.~\ref{fig4}(d) and \ref{fig4}(e).
As anticipated, the evolution of Coulomb resonances of the CNT QD (shown in Fig.~\ref{fig4}(d)) follows a constant positive slope with a relative lever arm of $\rm{\alpha}_{REF,BG}\approx$1.25 and the occupation numbers (denoted in white letters) decrease with increasing $\rm{V}_{REF}$ and decreasing $\rm{V}_{BG}$. For the charge detector the observed patterns in Fig.~\ref{fig4}(e) show similarities to those of a multi-dot system. Further analysis of the transconductance $\rm{dG}_{CD}/\rm{dV}_{REF}$ reveals at least three lines with negative slopes. These lines show that the transport through the GNR is governed by a series of quantum dots where each of the lines can be attributed to single QDs located at different positions along the nanoribbon. The extracted lever arms of the visible conductance resonances are $\rm{\alpha}_{REF,BG}^{(1)}$=$-0.43$, $\rm{\alpha}_{REF,BG}^{(2)}$=$-0.31$ and $\rm{\alpha}_{ REF,BG}^{(3)}$=$-0.12$ as indicated in Fig.~\ref{fig4}(f). Additionally the plot exhibits clearly visible features (see white dotted line) with in this case positive slopes with their positions matching perfectly to those in Fig.~\ref{fig4}(d). Consequently they can be associated with charging events in the CNT QD which are detected by the quantum dots situated in the graphene nanoribbon. Detection lines are even visible where the current through the dot is too low to be measured directly (see white arrows in Figs.~\ref{fig4}(d) and (f)).
In order to further improve the sensitivity, the detector could potentially be included in a RF-readout setup. The charge sensitivities reported for this type of setups are in the order of $~10^{-4} \rm{e}/\sqrt{\rm{Hz}}$ to $~10^{-5} \rm{e}/\sqrt{\rm{Hz}}$  \cite{got08,bie06}. While a quantitative comparison of the presented dc-readout technique with other dc-charge sensors on carbon nanotube QDs \cite{chu09, zho12} can not be provided (no sensitivities are given), a qualitative comparison of the data shows that the graphene charge detector yields at least equal visibility of the charge states.
\newline
Finally, we discuss a carbon nanotube device where both metal leads (source and drain) are substituted by two graphene flakes. The device structure has been discussed earlier (see fabrication section and Fig.~\ref{fig1}(e)) and a close up of the final device is shown as inset in Fig.~\ref{fig5}(a). For measuring the conductance of the nanotube, Cr/Au metal contacts are deposited on each graphene flake by an EBL step followed by metalization and lift-off. The metal contacts are designed in a way that they do not touch the carbon nanotube. As a first important observation we see that current flows from one graphene flake to the other through the carbon nanotube. Moreover, we do not observe a $\rm{V}_{BG}$-regime where the current is fully suppressed (not shown) which suggests the presence of a metallic nanotube. In Fig.~\ref{fig5}(b) we show the current $\rm{I}_{CNT}$ in dependence of the voltage $\rm{V}_{SG}$ applied to a metallic side gate which was deposited next to the nanotube as indicated by the blue area in Fig.~\ref{fig5}(a) and allows to locally gate the central segment of the carbon nanotube. The current $\rm{I}_{CNT}$ exhibits periodic Coulomb oscillations. In Fig.~\ref{fig5}(c) we plot the current through the nanotube $\rm{I}_{CNT}$ on a logarithmic scale as function of the bias voltage $\rm{V}_{BIAS}$ and $\rm{V}_{SG}$ for a fixed value of $\rm{V}_{BG}$=35.15 V. The measurement exhibits clear diamond shaped patterns which can be attributed to Coulomb diamonds. From the addition energies of 1-1.5 $\rm{meV}$ we can estimate the quantum dot length of 0.9-1.5 $\rm{\mu m}$ \cite{boc97}, which is in good agreement with the 1.2 $\rm{\mu m}$ spacing between the two graphene leads. This leads to the conclusion that the quantum dot in the nanotube extends over the full length set by the distance of the graphene leads. Moreover, the observation of Coulomb blockade oscillations and peaks manifests the presence of tunneling barriers at the interface of both carbon allotropes. From the maxima of the Coulomb peaks and assuming similar rates for tunneling in and out of the dot we extract tunneling rates in the low GHz regime. The origin of the observed tunneling barriers is not fully understood at the present state. Since graphene is a semi-metal and the investigated carbon nanotube is a metallic carbon nanotube, it is believed that the tunneling barriers form similarly to those metallic-CNT devices with metal contacts. Here, tunneling barriers may arise from imperfect electrical contacts to the nanotube \cite{woo02}.
\begin{figure}[tb]\centering
\includegraphics[draft=false,keepaspectratio=true,clip,%
                   width=1\linewidth]%
                   {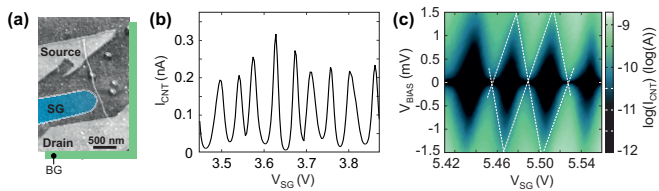}
\caption[FIG5]{(color online)
(a) SFM image of the investigated device (see also Fig.~\ref{fig1}(e)).
(b) Current $\rm{I}_{CNT}$ through the nanotube as function of the voltage $\rm{V}_{SG}$ applied to a metal electrode placed roughly 50~nm from the nanotube (see blue area in (a)) for a fixed back gate voltage $\rm{V}_{BG}$=35~V and $\rm{V}_{BIAS}$=0.3 mV.
(c) Current $\rm{I}_{CNT}$ through the nanotube on a logarithmic scale as function of $\rm{V}_{BIAS}$ and $\rm{V}_{SG}$. $\rm{V}_{BG}$ is constant at 35.15 V.}
\label{fig5}
\end{figure}
\newline
In summary, we present the fabrication and characterization of carbon nanotube-graphene hybrid devices. We show the example of a structure where a nanotube quantum dot is capacitively coupled to a graphene nanoribbon. Sharp resonances in the graphene nanoribbon conductance give rise to clear charge detection signals with an upper estimate for the charge sensitivity of $~10^{-3} \rm{e}/\sqrt{\rm{Hz}}$. This kind of graphene charge detector represents a third charge detection scheme for CNT QDs in addition to the existing strategies based on metal single electron transistors \cite{got08,bie06} and capacitively coupled carbon nanotube segments \cite{chu09, zho12}. The presented charge detector is rather easy to fabricate. Its sensitivity can be further improved by including it into a RF-circuit. In addition, we study a device where two carbon allotropes (graphene and CNT) are electrically coupled. We find that the interfaces between the graphene and the carbon nanotube resembles tunneling barriers. This device represents an important preliminary step towards experiments investigating the quantum phase transition from screened to unscreened impurity moments. Both results open the road to more sophisticated devices which are entirely fabricated out of carbon nanostructures and exploit the different advantages of these promising materials.

{Acknowledgment ---}
The authors wish to thank L.~Durrer and C.~Hierold for their support and help
on the growth of the CNT samples.
We thank A.~Steffen, R.~Lehmann and J.~Mohr for the help on the sample fabrication.
Discussions with
F.~Haupt, J.~Splettst\"osser, D. Schuricht and M.~Wegewijs
 and support by the JARA Seed Fund and the DFG
(SPP-1459 and FOR-912) are gratefully acknowledged.

\end{document}